\documentclass[journal=jacsat,manuscript=article]{achemso}

\usepackage[T1]{fontenc} 
\usepackage{upgreek} 
\usepackage[utf8]{inputenc} 
\usepackage{graphicx} 
\usepackage{bm} 
\usepackage{amsmath} 
\usepackage{caption} 
\newcommand{\un}[2]{#1\,\mathrm{#2}} 

\title{A light induced metastable magnetic texture uncovered by in-situ Lorentz microscopy}

\author{Tim Eggebrecht}
\affiliation{I. Physical Institute, Georg-August-University, G\"ottingen, Germany}
\altaffiliation{These authors contributed equally to this work.}

\author{Marcel M\"oller}
\affiliation{IV. Physical Institute, Georg-August-University, G\"ottingen, Germany}
\altaffiliation{These authors contributed equally to this work.}

\author{J. Gregor Gatzmann}
\affiliation{IV. Physical Institute, Georg-August-University, G\"ottingen, Germany}

\author{Nara Rubiano da Silva}
\affiliation{IV. Physical Institute, Georg-August-University, G\"ottingen, Germany}

\author{Armin Feist}
\affiliation{IV. Physical Institute, Georg-August-University, G\"ottingen, Germany}

\author{Ulrike Martens}
\affiliation{Interface and Surface Physics, Ernst-Moritz-Arndt-University, Greifswald, Germany}

\author{Henning Ulrichs}
\affiliation{I. Physical Institute, Georg-August-University, G\"ottingen, Germany}

\author{Markus M\"unzenberg}
\affiliation{Interface and Surface Physics, Ernst-Moritz-Arndt-University, Greifswald, Germany}

\author{Claus Ropers}
\affiliation{IV. Physical Institute, Georg-August-University, G\"ottingen, Germany}

\author{Sascha Sch\"afer}
\affiliation{IV. Physical Institute, Georg-August-University, G\"ottingen, Germany}
\email{schaefer@ph4.physik.uni-goettingen.de}

\begin{document}
	
\begin{abstract}
Magnetic topological defects, including vortices and Skyrmions, can be stabilized as equilibrium structures by tuning intrinsic magnetic interactions and stray field geometries. Here, employing rapid quench conditions, we report the observation of a light-induced metastable magnetic texture, which consists of a dense nanoscale network of vortices and antivortices and exhibits glass-like properties. Our results highlight the emergence of complex ordering regimes in optically driven magnetic systems, opening up new pathways to harness ultrafast far-from-equilibrium relaxation processes.  
\end{abstract}

Topologically protected magnetic defects cannot be continuously transformed into a defect-free state and therefore may be viewed as quasi-particles, largely robust against thermal perturbation \cite{bib:braun12}. Depending on the magnetic anisotropy of the structure, magnetic defects of different topology, such as vortices and skyrmions, may exist, often stabilized in tailored nano-structures \cite{bib:shinjo00,bib:wachowiak02,bib:uhlig05,bib:szary10,bib:pollard12} or by intrinsic interactions \cite{bib:rossler06, bib:nagaosa13}. For the latter case, as a prominent example, the chiral Dzyaloshinskii-Moriya interaction induces ordered Skyrmion lattices in helimagnetic materials, such as MnSi \cite{bib:muehlbauer09}, $\textrm{Fe}_{1-x}\textrm{Co}_x\textrm{Si}$ \cite{bib:yu10}, or in single-atomic iron layers on an iridium surface \cite{bib:heinze11}.

As expected for robust quasi particles, magnetic defects can be manipulated by external perturbations, including (spin-polarized) currents \cite{bib:klaui05,bib:yamada07,bib:jiang15}, magnetic fields \cite{bib:vanwaeyenberge06,bib:buttner15}, thermal gradients \cite{bib:bauer12,bib:schlickeiser14}, and optical excitations \cite{bib:finazzi13}, but their temporal evolution remains to be constrained by topological invariants. For example, for vortices stabilized in magnetic nanodiscs, the reversal of the vortex core polarization upon radiofrequency excitation was observed to occur only over a sequence of vortex-antivortex generation/annihilation events, instead of the topologically forbidden direct flip of the core polarization \cite{bib:hertel06, bib:vanwaeyenberge06,bib:gomez11,bib:kammerer11}. 

Despite their intrinsic stability, topological defects can be created or destroyed at structural inhomogeneities, including particle boundaries \cite{bib:buttner15} or channel constrictions \cite{bib:jiang15}. More fundamentally, within a homogeneous film, topological defects may be formed in pairs of opposite topological character, such as vortices and antivortices, a process reminiscent of particle-antiparticle creation. Topological defect pair generation is particularly effective when crossing a second-order phase transition at fast quench rates, and was described to follow a universal behavior captured within the framework of the Kibble-Zurek mechanism (KZM) \cite{bib:kibble76,bib:zurek85,bib:lin14}.  For non-magnetic degrees-of-freedom, quench-induced defect pair generation was experimentally demonstrated in several solid-state systems, including nematic liquid crystals \cite{bib:chuang91}, ferroelectric materials \cite{bib:griffin12,bib:lin14} and in superfluidic $^3$He \cite{bib:ruutu96,bib:bauerle96}, as well as in confined atomic neutral and ionized gases \cite{bib:ulm13}. In contrast, in magnetic systems, KZM-like defect pair generation has remained elusive, and is only expected to be observed at large quench rates.

In this work, we demonstrate the formation of a metastable magnetic vortex-antivortex texture in a thin ferromagnetic iron layer after quenching from the paramagnetic state on picosecond time scales. Femtosecond optical excitation with a single laser pulse demagnetizes a micrometer-sized region of the iron film, and subsequent thermal quenching due to substrate results in a dense nanoscale network of localized magnetic vortices and antivortices. Spatial correlations within the network structure are mapped in-situ by Lorentz microscopy, exhibiting glass-like properties.

\begin{figure}
	\includegraphics[width=0.58\linewidth]{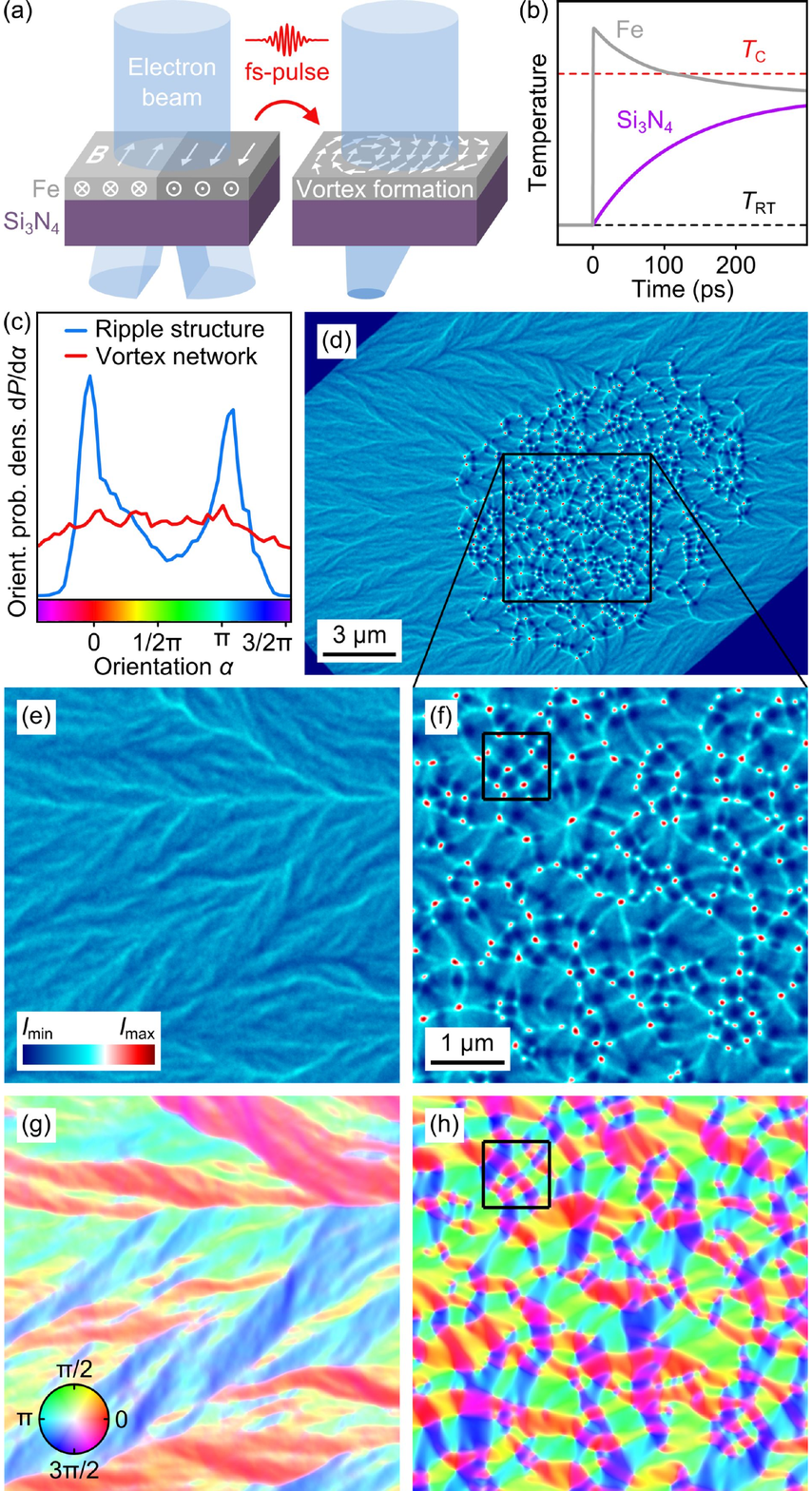}
	\caption{Optically induced magnetic vortex-antivortex network. \textbf{(a)} The in-plane magnetization of an iron thin film is imaged by Lorentz microscopy employing in-situ optical excitation. \textbf{(b)} Simulated temperature transient after optical excitation in the iron thin film and the silicon nitride substrate ($T_\text{C} = \un{1041}{K}$: Curie temperature of iron, $T_\text{RT} = \un{300}{K}$: room temperature, fluence $F=\un{10}{mJ/cm^2}$)  \textbf{(c)} Orientation probability of the in-plane magnetization shown in (g) and (h), demonstrating the transition from an oriented magnetic ripple structure into a largely isotropic vortex network. \textbf{(d-f)} Electron micrographs with Lorentz contrast before (e) and after (d,f) applying a single femtosecond laser pulse (optical fluence $F=\un{12.7}{mJ/cm^2}$). (The dark blue regions in the corners of (d) correspond to the opaque silicon support frame) \textbf{(g,h)} Reconstructed in-plane magnetization for the Lorentz images shown in (e) and (f), respectively. Color saturation and hue signify the normalized magnitude and direction of the local magnetization (see color wheel in (g)). Close-ups of the marked areas in (f) and (h) are shown in Fig. 2. (Equal length scales are chosen in (e-h). Electron intensity and magnetization were normalized to the maximum values in (f) and (h), respectively.)}
	\label{fig:expSetup}	
\end{figure}

In our experiments, iron/silicon-nitride bilayers were prepared by depositing a polycrystalline iron thin film ($\un{10}{nm}$ thickness, about $\un{25}{nm}$ grain size) onto silicon nitride membranes ($\un{20}{nm}$ thickness, $\un{28}{\upmu m}\times\un{17}{\upmu m}$ membrane area) using electron beam evaporation in an ultrahigh vacuum environment. To map the optically induced magnetization changes in the bilayer system, we perform Lorentz microscopy in a transmission electron microscope, which was modified to allow for \emph{in-situ} laser excitation. Out-of-focus conditions are utilized to acquire images with a contrast sensitive to the transverse sample magnetization \cite{bib:petford12}. High throughput single-shot imaging allows us to microscopically investigate irreversible magnetization changes and to acquire statistical properties of the optically induced changes. Magnetization changes are initiated by single ultrashort laser pulses ($\un{800}{nm}$ center wavelength, $\un{150}{fs}$ pulse duration, $\un{45}{\upmu m}$ focal spot size), and the magnetic structure is imaged after each single laser pulse (cf. Fig.~1(a)). Laser excitation at high pulse fluences results in an ultrafast heating of the iron film above its Curie temperature, followed by a rapid temperature quench due to the adjacent silicon nitride substrate. The estimated temperature transients of both layers in the center of the laser profile are shown in Fig.~1(b) for a peak optical fluence of $F=\un{10}{mJ/cm^2}$ (see Supplemental Material for details). Notably, in this configuration, cooling speeds larger than $10^{12}$~K/s are achieved in the iron layer.

Figures 1(e) and (d,f) display Lorentz electron micrographs of the magnetic bilayer before and after applying a single femtosecond laser pulse, respectively (see also Supplemental Material, Movie S1). For an optical excitation above a well-defined fluence threshold of $F=\un{11.5}{mJ/cm^2}$, the initial magnetic structure within a sharply demarcated area at the center of the optical focus undergoes a remarkable change (Fig. 1(d)), from a weak meander-like contrast (Fig. 1(e)) to an image with high-contrast hot and cold spots (Fig.~1(f)). Optical pulses below this threshold only result in increased domain wall mobility with no large-scale magnetic reorganization (see Supplemental Material and Movie S2). 

From the Lorentz micrographs, we reconstruct the underlying in-plane magnetization for both cases (Fig.~1(g,h)), using a transport-of-intensity approach \cite{bib:petford12, bib:volkov02} (see Supplemental Material). Prior to optical excitation, the magnetic texture in the iron thin film shows a ripple structure\cite{bib:methfessel61} with two preferred antiparallel magnetization directions (cf. Fig.~1(c), blue curve). After optical excitation, the magnetic structure exhibits no preferred orientation (Fig.~1(c), red curve) but instead is characterized by a large number of nanoscale domains. 

Figs.~2(a,c) display the Lorentz micrograph and the reconstructed in-plane magnetization of a small sample area (marked in Figs.~1(f,h)), respectively. Vortices and antivortices are formed at domain junctions, yielding a dense defect network. At each magnetic vortex (open/dotted white discs), the in-plane magnetization curls in a clockwise/counterclockwise direction around a central point leading to a local focussing/defocussing of the incident electron beam due to the Lorentz force. Thereby, bright/dark contrasts are generated in the Lorentz images \cite{bib:degraef03}. Between vortices with equal curling direction, antivortices are located (red circles), which are discernible as saddle points in image intensity.

\begin{figure}
	\includegraphics[width=0.58\linewidth]{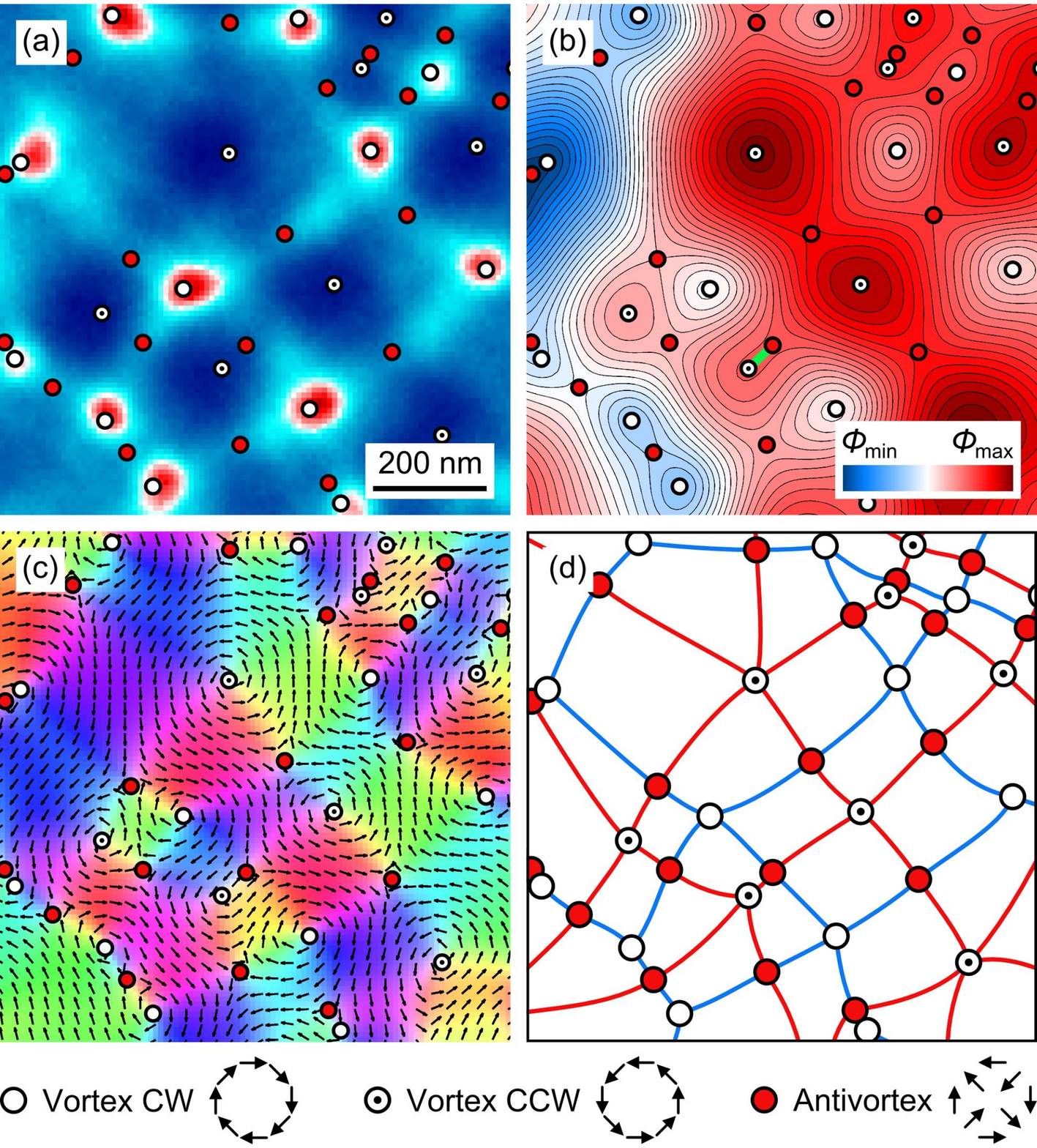}
	\caption{Magnetic network structure. \textbf{(a,c)} Lorentz micrograph (a) and corresponding reconstructed in-plane magnetization (c) for the sample area indicated in Fig.~1(f,h). The positions of clockwise (CW) and counterclockwise (CCW) rotating vortices and antivortices were retrieved from the magnetization maps and are indicated as white and red circles, respectively (color scales equal to Fig.~1(f) and (h), respectively; vectors in (c) indicate magnetization direction). \textbf{(b)} Reconstructed local phase which is imprinted by the magnetic network onto the imaging electron wave. Phase minima/maxima result in an electron beam focussing/defocussing leading to image intensity maxima/minima. Anti-vortices are located at saddle points of the phase surface. Green bar indicates a closeby vortex-antivortex pair. \textbf{(d)} Topology of vortex-antivortex network constructed from the phase map in (b) following the steepest ascent and descent paths, which start at each phase saddle point.}
	\label{fig:vortexZoom}
\end{figure}

In Fig. 2(b), the spatial distribution of topological defect types within the network is rationalized by considering the local phase imprinted on the transmitted electron microscope beam. Here, clockwise and counterclockwise vortices correspond to maxima and minima in the phase map, respectively, whereas antivortices can be seen as saddle points. A path from one phase basin to the next can be naturally chosen to include a saddle point, resulting in an interwoven network with alternating vortex and antivortex nodes (Fig. 2(d)). 

A more detailed analysis of the average distances between these defects is obtained from radial pair-correlation functions (cf. Fig. 3(d) at $F=\un{14.4}{mJ/cm^2}$), involving equally rotating vortices ($\varrho_{vv}$, green curve), vortices and antivortices ($\varrho_{av}$, red curve), and vortices with different rotation directions ($\varrho_{vv^*}$, blue curve). $\varrho_{av}$ and $\varrho_{vv^*}$ exhibit pronounced maxima (at $\un{80}{nm}$ and $\un{150}{nm}$, respectively, demonstrating short range order between neighboring nodes in the defect network. $\varrho_{vv}$ displays only a very weak maximum around $\un{200}{nm}$. In addition, we find a minimal node distance of $\un{50-100}{nm}$, leading to an excluded area around each defect. We note that the magnetic configuration arising from a pair of closeby vortex and antivortex defects can be locally  transformed into a defect-free state \cite{bib:chaikin00}, resulting in hairpin-like equal-phase curves as indicated in Fig.~2(b) for one connected vortex-antivortex pair (green bar). Similar to atomic pair correlation functions of glasses and liquids, no spatial network ordering is observed beyond nearest neighbor distances. Furthermore, for subsequent above-threshold laser pulses, the position of vortices and antivortices on the sample is randomly distributed within the center of the excited sample region (Supplemental Material, Fig.~S5), highlighting that pinning sites \cite{bib:uhlig05} on length scales larger than the grain size of the polycrystalline iron film only play a minor role.

In Fig.~3(a), the number of laser-generated vortices and antivortices is displayed for different optical fluences, observing a linear increase above a well-defined fluence threshold. Taking into account the fluence-dependent network area yields a saturation behaviour of the defect density at high fluences (cf. 3(e,f)). Vortices and antivortices are topological defects which can be characterized by a winding number $w=1/2\uppi \oint\bm{\nabla}\alpha\cdot\text{d}\bm{s}$ yielding $w=\pm 1$ for vortices and antivortices, respectively ($\alpha$: local orientation of in-plane magnetization vector) \cite{bib:chaikin00}. The integral extends over an arbitrary contour containing the magnetic singularities, and thereby locally confined magnetization changes conserve the total winding number in the affected area. Consequentially, as observed here, an equal number of laser-generated vortices and antivortices is expected after optical excitation.

\begin{figure}
	\includegraphics[width=0.58\linewidth]{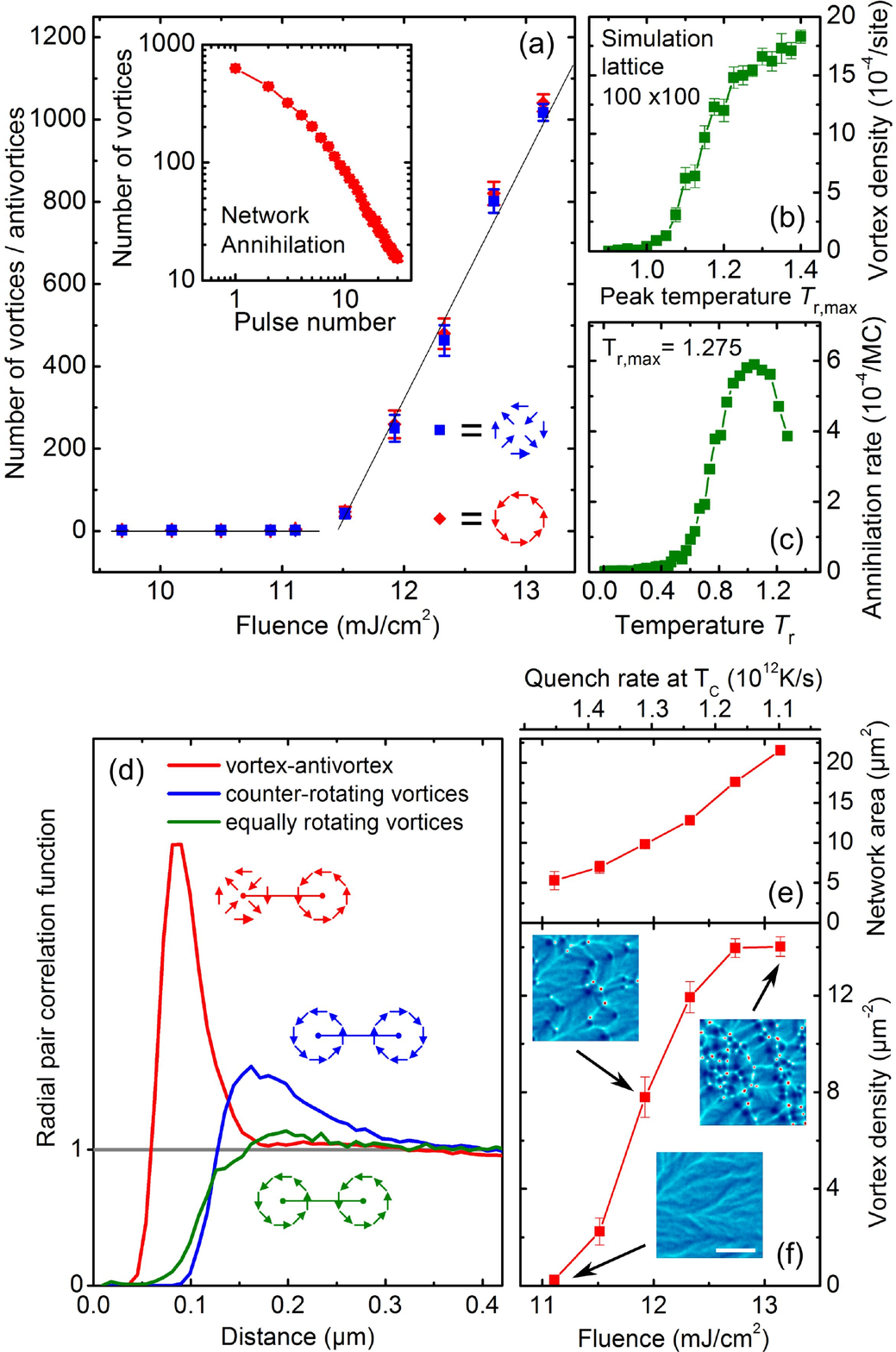}
	\caption{Fluence-dependencies and spatial network dimensions. \textbf{(a)} Experimentally observed total number of vortices and antivortices after applying single laser pulses with a varying optical fluence $F$, exhibiting a well-defined fluence threshold of about $\un{11.5}{mJ/cm^2}$. Inset: Optically-triggered defect annihilation below fluence threshold. Starting with a vortex-antivortex network generated above threshold, subsequent optical pulses at $F=\un{10.8}{mJ/cm^2}$ restore the magnetic ripple structure. \textbf{(b)} Monte-Carlo simulation of the number of remaining vortices after rapid quench from a reduced peak temperature $T_\text{r,max}$. \textbf{(c)} Simulated vortex-antivortex annhihilation rate (per Monte-Carlo step (MC)) at reduced temperatures $T_\text{r}$ during quench from $T_\text{r,max}=1.275$. A substantial number of annhihilation events occur far below the threshold temperature ($T_\text{r,vortex}=1.05$). \textbf{(d)} Experimental radial pair correlation function between the position of vortices and antivortices ($\varrho_{va}$, red), and equal- ($\varrho_{vv}$, green) and counter-rotating ($\varrho_{vv^*}$, blue) vortices. \textbf{(e)} Fluence-dependent area of the vortex-antivortex network ($A=\uppi\sigma_x\sigma_y$), defined by the standard deviations $\sigma_{x,y}$ of vortex positions in both spatial directions. \textbf{(f)} Vortex density depending on the excitation fluence, reaching a maximum value of about $\un{14}{vortices/\upmu m^2}$ (scale bar: $\un{1}{\upmu m}$). Insets show representative electron micrographs at the indicated fluences. Estimated quench rate at $T_\text{C}$ is shown for the employed fluence range.}
	\label{fig:fluence}
\end{figure}

The spontaneous generation of vortices and antivortices by light in a homogeneous magnetic thin film may at first seem surprising. In localized nanoparticles, vortex structures are typically found as the lowest-energy configuration, stabilized by stray field minimization \cite{bib:shinjo00,bib:shigeto02,bib:wachowiak02,bib:uhlig05,bib:szary10}, whereas for homogeneous thin films, as studied here, vortex structures are excited states. Nevertheless, we find the laser-induced vortex-antivortex network to be stable at room temperature, and no relaxation to the equilibrium magnetic ripple structure is observed over a time span of several months. However, in a creep-like process similar to structrural glasses, vortex-antivortex annihilation processes can be facilitated by applying optical pulses with a fluence slightly below the threshold for vortex-antivortex generation, evidencing the glassy character of the light-induced vortex-antivortex texture. The decrease of the number of vortices for subsequent intermediate intensity pulses is quantified in the inset of Fig.~3(a), with the equilibrium magnetic ripple pattern completely regained after several tens of optical pulses (see Supplemental Material, Movie S3, for a sequence of Lorentz images). The lack of long-range order and the appearance of creep are hallmarks of a glassy state, which differentiates the light-induced magnetic vortex-antivortex network from a super-cooled high-temperature equilibrium magnetic phase \cite{bib:oike16}, and points to strong analogies to structural glasses as well as phase-change materials \cite{bib:wuttig07,bib:tuma16}.  

In the following, we describe in more detail the generation mechanism leading to the formation of the vortex-antivortex network. After optical excitation, the magnetic structure of the iron film will be strongly altered when its peak spin temperature exceeds the Curie temperature. At the ferromagnetic phase transition, the averaged local magnetization approaches zero, leading to substantially decreased magnetic dipolar couplings and shape anisotropy energy, and to a divergent magnetic response time upon external perturbations \cite{bib:chubykalo06}. Slowed-down dynamics at second-order phase transitions were invoked first by Kibble \cite{bib:kibble76} in a cosmological context and later for condensed matter systems by Zurek \cite{bib:zurek85} to predict the appearance of topologically protected configurations when crossing the phase transition at finite cooling rates. In particular, the order parameter of a system freezes out in a narrow temperature interval around the phase transition, and features of the high temperature phase are introduced into the newly formed symmetry-broken phase as topological defect states. In agreement with this description, vortex-antivortex generation is experimentally observed for optical fluences above $F=\un{11.5}{mJ/cm^2}$ (cf. Fig.~3(a), estimated peak temperature at threshold: ${\sim}\un{1400}{K}$). 

The average node distance within the vortex-antivortex network of about $\un{100}{nm}$ (Fig.~3(d)) reflects the coherence length $l_\text{c}$ of the newly formed magnetic texture, and, within the KZM, is expected to increase with increasing quench rate \cite{bib:zurek85}. Such a behavior was, e.g., found in the case of a ferroelectric phase transition \cite{bib:lin14} for quench rates in the range of $\un{10-10^4}{K/h}$). The ultrafast quenching scheme employed here with cooling speeds larger than $\un{10^{12}}{K/s}$ is not readily adaptable to larger changes in the quench rate. Specifically, we estimate that cooling speeds at the phase transition temperature vary by only a factor of 1.3 in the investigated fluence interval, with the highest quench rates close to the threshold fluence. Contrary to the predictions of the KZM, the largest vortex density is experimentally observed at the highest fluence values, i.e. at the smallest quench rates at $T_\text{C}$ (Fig.~3(f)), indicating that vortex-antivortex annihilation events occuring after the phase transition significantly alter the observed magnetic texture.

To obtain further insights into the creation and annihilation dynamics of topological defects, we conducted numerical simulations of the dynamics of in-plane oriented classical spins. Within a Metropolis Monte-Carlo 2D XY model \cite{bib:chaikin00}, we consider a quadratic spin lattice, coupled by nearest-neighbor exchange interactions and connected to a heat bath of variable temperature (see Supplemental Material). Starting with a thermal spin distribution at an elevated temperature $T_\text{r}$ (reduced temperature scale normalized to the exchange energy), the system is cooled down with a fixed quench rate, and the number of remaining vortices is extracted. As shown in Fig. 3(b), vortices only remain after cool down, if the reduced temperature exceeds a well-defined temperature threshold $T_\text{r,vortex}=1.05$. Despite the highly simplified theoretical model, the threshold behavior and the dependence of the vortex density on peak temperature closely follow the experimentally observed behavior. Furthermore, in the numerical simulation, the number of defects substantially decreases during cool-down by pronounced vortex-antivortex annihilation, occuring even at reduced temperatures far below $T_\text{r,vortex}$ (Fig. 3(c)). After cool-down, the vortex number is reduced to about 5\% of its initial value at $T_\text{r}=1.05$, leading to a five-fold increase in nearest-neighbor distance between defects. These results further indicate that, after vortex-antivortex generation for peak temperatures close to $T_\text{C}$, the observable defect density may crucially depend on the subsequent dynamics during cool-down to room temperature, including coarsening processes \cite{bib:biroli10} with potential contributions from the network boundary. In particular, for mobile topological defects, vortex-antivortex annihilation may occur, reducing the final defect density, and increasing the characteristic nearest neighbor distances within the vortex-antivortex network. Consequentially, the number of observed vortices and antivortices will show a more complex dependence on the overall cooling rate, in contrast to a Kibble-Zurek-type scaling with the quench rate at the phase transition.  

Finally, we note that the excitation and quenching conditions necessary to induce a nanoscale vortex antivortex network are similar to the ones typically employed in helicity-dependent all-optical switching \cite{bib:stanciu07,bib:lambert14}, and metastable states as observed here may generally play an important role in the manipulation of magnetic structures by intense light.

In conclusion, we demonstrated the optically triggered formation of a magnetic vortex-antivortex texture and analyzed its structural properties using in-situ Lorentz microscopy. Rapid quenching preserves a high density of topological defects, which arranges in an extended network structure. This approach may give access to a host of metastable configurations in various magnetic materials, and provides novel opportunities to unravel nanoscale ordering mechanisms far-from-equilibrium, complementary to emerging approaches in ultrafast transmission electron microscopy \cite{bib:zewail10,bib:feist15}.

\begin{acknowledgement}

We gratefully acknowledge fruitful discussions with K. Samwer and M. Plenio. Financial support was provided by the Deutsche Forschungsgemeinschaft (DFG-SFB 1073/projects A05 and B01), the Volks\-wa\-gen\-Stif\-tung, and the Lower Saxony Ministry of Science and Culture.

\end{acknowledgement}

\bibliography{literature} 

\end{document}